\documentclass[twoside, epsfig]{article}

\input oejv.sty

\begin{document}

\OEJVhead{March 2015}
\OEJVtitle{Possible candidates for multiple occurrence}\vspace{-0.4cm}
\OEJVtitle{of variable stars in the VSX catalogue}
\OEJVauth{Li\v{s}ka, J.$^{1,2}$; Skarka, M.$^{1,2}$; Auer, R.~F.$^{2,3}$; Prudil, Z.$^{1}$; and Jur\'{a}\v{n}ov\'{a}, A.$^{1}$}

\OEJVinst{Department of Theoretical Physics and Astrophysics, Masaryk University, Kotl\'{a}\v{r}sk\'{a} 2, 611 37 Brno,\\ Czech Republic; {\tt \href{mailto:jiriliska@post.cz}{jiriliska@post.cz}}}
\OEJVinst{Variable Star and Exoplanet Section of the Czech Astronomical Society, Vset\'{i}nsk\'{a} 941/78,\\ 757 01 Vala\v{s}sk\'{e} Mezi\v{r}\'{i}\v{c}\'{i}, Czech Republic}
\OEJVinst{South-Moravian-Observatory, 664 71 Chud\v{c}ice 273, Czech Republic}

\OEJVabstract{A paper about variable stars with possible multiple occurrence in the VSX catalogue is presented. Our main criteria for identification of such duplicities were the angular distance among stars (below 1 arcmin) and close periods of objects. In our approach, we also considered double or half values of periods to reveal possible misclassification among stars with similar light curve shapes. The probability of false identification is expressed by the parameter $R$ giving the relative difference between periods. We found 1487 pairs of stars in angular distance lower than 1\,arcmin with period difference $R$ lower than 0.1\,\%, which are high-probable candidates on duplicates. From this sample, 354 pairs have exactly the same periods $(R = 0.0\,\%)$ and should be considered as definite duplicates. The main contribution of certain duplicates comes from the Catalina Sky Survey (73 pairs have two names with CSS acronym) and from the BEST projects (71 pairs). Distribution of identified duplicates on the sky is not homogeneous but contains surprising depression in Galactic plane.}

\begintext

We present a by-product of a new project SERMON ({\it SEarch for Rr lyraes with MOdulatioN}) which focuses on searching for modulation in RR Lyrae stars\footnote{Details of the project and the main results will be published in near future \citep[][in prep.]{skarka2015}.}. During preprocessing of the data for chosen RR Lyrae stars we noticed that there are several suspicious objects with very similar position and period among our targets. This finding initiated an analysis of the entire International Variable Star Index \citep[VSX, version from February 9, 2015,][]{watson2006}\footnote{http://www.aavso.org/vsx/index.php} containing 325061 records (including suspected and non-variable objects).

Since our initial finding suggested that the misclassification comes mainly from the large sky surveys like ASAS\footnote{http://www.astrouw.edu.pl/asas/} \citep[e.g.][]{pojmanski2002}, NSVS\footnote{http://skydot.lanl.gov/nsvs/nsvs.php} \citep{wozniak2004} or Catalina Surveys\footnote{http://nesssi.cacr.caltech.edu/DataRelease/} \citep[CSS or CRTS,][]{drake2009} we assume that the problem could mainly be caused by their typically small angular resolution (several arcsec/pixel) and therefore such troubles are expected to appear preferably in crowded stellar fields (Milky Way, Galactic bulge, stellar clusters, etc.) where the cross-identification can be difficult. Additional biases could be caused by bright stars and atmospheric conditions. Bad seeing will definitely influence the angular resolution, and because large sky surveys are unable to avoid overexposure, the light of very bright stars can influence photometry of nearby objects as well. Instrumental artefacts such as diffraction patterns or bad pixels are additional effects which could bring problems.

Thus we decided to identify stars which are closer than 1 arcmin on the plane of the sky and check whether they have close periods and eventually similar variable types. We did not perform detail astrometry or visual inspection of particular cases, therefore the only criteria considered in our analysis were the formerly noted.

Angular distance $r$ in arcsec between two objects was determined from their right ascensions $\alpha_{1}$, $\alpha_{2}$ and declinations $\delta_{1}$, $\delta_{2}$ given in VSX in degrees by equation
\begin{equation} \label{eqradius}
r = 3600\,\sqrt{(\alpha_{1}-\alpha_{2})^2\cos^2\left(\frac{\delta_{1}+\delta_{2}}{2}\right) + (\delta_{1}-\delta_{2})^2}.
\end{equation}

For a comparison whether two variables are in fact one single star, or a star misclassified as a nearby variable, we introduce parameter $R$ that represents relative difference between their periods $P_{1}$ and $P_{2}$ as 
\begin{equation} \label{eqratio}
R = \frac{|P_{1}-P_{2}|}{\displaystyle\left(\frac{P_{1}+P_{2}}{2}\right)} \times 100\,\%.
\end{equation}
The period-similarity criterion is the only usable since different surveys use different devices and filters and also stars itself may undergo significant brightness changes -- e.g. recurrent novae. Thus any information about brightness could be misleading. The VSX catalogue contains 264542 stars with known period (81.4\,\%)\footnote{97 stars have zero-period and were not analysed.}. Unfortunately we cannot provide any closer information about non-periodic variables which can cause significant bias between our results and real fraction of duplicates. 

If we apply $R$ parameter, the number of stars which have one or more stars in angular distance lower than 1\,arcmin (91315 objects) significantly reduces. For example, number of pairs from this set with $R<5$\,\%, and thus potential candidates, is 5549. However, estimation of appropriate $R$ is not straightforward. Many variable stars located in small area in dense stellar fields (Galactic bulge) can have similar periods, but are clearly different variables. In addition, due to similarity of light curves for some variable types (e.g. RRc and EW) when using poor quality data, these objects can be easily misclassified. Therefore we also checked the possibility that one object has double/half period of the other star\footnote{Number of pairs from this set including possibility of double/half value of period with $R<5$\,\% is 9151.}.

We consider that stars with $R<0.1$\,\% are almost certainly duplicates since this ratio translates to a difference between periods of 0.0006\,d ($\sim$ 1\,minute) in stars with periods about 0.6\,d and in period-difference of 0.7\,d in stars with period about 700-day. 1487 pairs comply this condition (Tab.~\ref{table1}, the whole table is only available on-line as a supporting material). Inside this group we found 354 pairs of stars with $R=0.0$\,\%, thus these stars are definitely duplicates. The rest of stars should be considered as possible duplicates with high probability. We point out that there are also stars included several times (pairs with different objects), because they create multiple-occurrence objects.

\begin{table}[thbp]
\setlength{\tabcolsep}{2.5pt}
\tiny
\centering
\caption{A list of stars with probable double occurrence in the VSX catalogue. Parameter $R$ is relative difference of periods, $r$ is mutual angular distance between both positions, $\alpha_{1}$, $\delta_{1}$, $\alpha_{2}$, $\delta_{2}$ are equatorial coordinates (J2000.0) for both objects,  $P_{1}$ and $P_{2}$ are their periods and VAR$_{1}$ and VAR$_{2}$ are variability types of candidates.}
\begin{tabular}{l l c c c c c c c c c c l}\\
\hline
Name 1 & Name 2 & $R$	& $r$     & $\alpha_{1}$ & $\delta_{1}$ & $\alpha_{2}$ & $\delta_{2}$ & $P_{1}$ & $P_{2}$ & VAR$_{1}$ & VAR$_{2}$\\
       &        & [\%]	& [arcsec]& [$\deg$]     & [$\deg$]     & [$\deg$]     & [$\deg$]     & [d]     & [d]     &           & \\
\hline\hline\\
GM And                            &   V0467 And                        & 0.005872 & 38.1410 & 0.01521 & 35.36286  & 0.02721 & 35.36692  & 0.7067585 & 0.3534   &   RRAB  &   EW	    \\
CSS\_J001145.6+181545             &   CSS\_J001145.7+181545            & 0.001344 & 1.4644  & 2.94029 & 18.26264  & 2.94071 & 18.26256  & 0.297666  & 0.297662 &   EW    &   EW     \\
CSS\_J001228.5+274432             &   CSS\_J001228.7+274431            & 0.000000 & 2.5535  & 3.11904 & 27.74231  & 3.11979 & 27.74206  & 0.23091   & 0.23091  &   EW    &   EW     \\
ASAS J001231-1402.1               &   CSS\_J001231.6-140208            & 0.005822 & 9.3048  & 3.12917 & -14.03500 & 3.13171 & -14.03578 & 0.326337  & 0.326318 &   RRC   &   RRC    \\
CSS\_J001610.2+273430             &   NSVS 6296844                     & 0.007069 & 11.1860 & 4.04283 & 27.57525  & 4.04480 & 27.57268  & 0.325347  & 0.32537  &   EW    &   EW     \\
CSS\_J001701.6+165938             &   ASAS J001702+1659.6              & 0.001061 & 5.9996  & 4.25679 & 16.99411  & 4.25833 & 16.99333  & 0.376966  & 0.376962 &   EW    &   ESD|EC \\
& & & & & \textbf{\vdots}& & & & & & \\
\hline\\
\end{tabular}\label{table1}
\end{table}

\begin{figure}[t!]
\centering
\includegraphics[width=12cm]{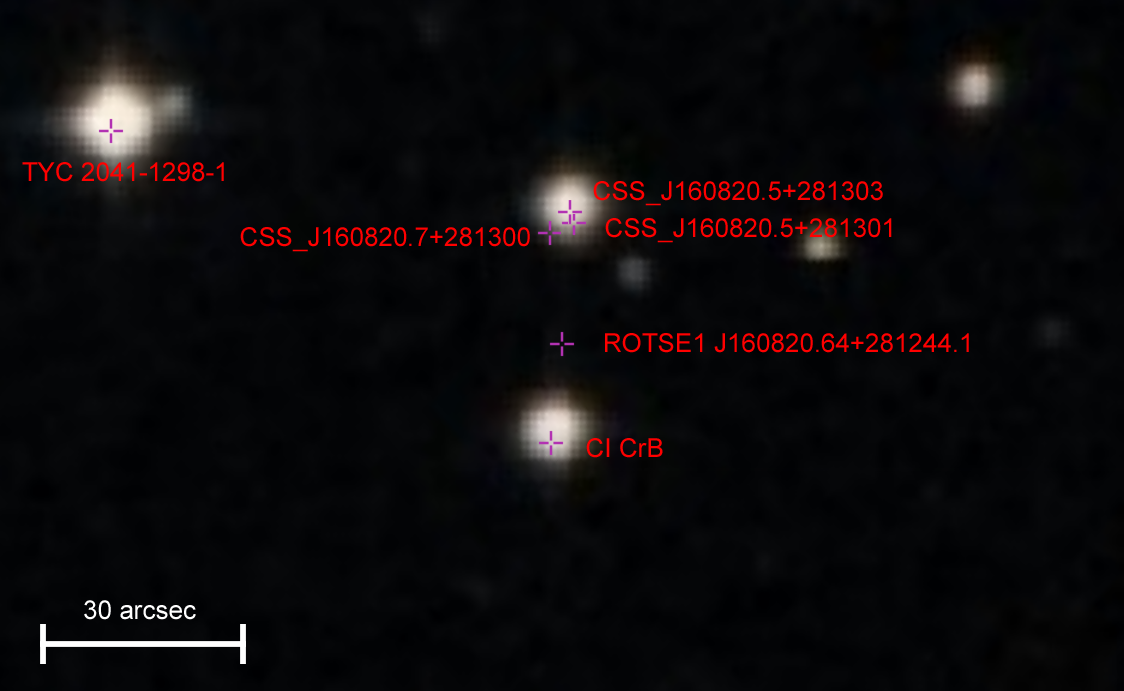} 
\caption{Position of CI CrB, ROTSE1 J160820.64+281244.1, and 3 almost identical objects with CSS names adopted from AladinLite view.}
\label{fig1}
\end{figure}

This is the case of CI CrB which was discovered by ROTSE-1 \citep{akerlof2000} as a $\delta$ Sct variable with a period of 0.17503799\,d. The Simbad database known only its name ROTSE1 J160820.64+281244.1. Later this star was reclassified as EW type by \citet{jin2004}. Nevertheless, CSS survey \citep{drake2014} found other 3 variables of EW types with almost identical periods in its close vicinity (Tab.~\ref{table2}). The position of CI~CrB corresponds to one star in the field, three objects with CSS names are concentrated in position of another star in the field and object ROTSE1 J160820.64+281244.1\footnote{Simbad coordinates for ROTSE1 J160820.64+281244.1 are $\alpha=16^{\rm h}08^{\rm m}20.643^{\rm s}$, $\delta=$+28$^{\circ}$12$'$44.12$''$ (J2000.0).} is located between CI CrB and CSS objects in a place without any star (Fig.~\ref{fig1}). 

\begin{table}[h]
\setlength{\tabcolsep}{4pt}
\centering
\caption{Nearby objects of CI CrB. The columns are identical as in Tab.~\ref{table1} ($R$ and $r$ parameters are relative to CI CrB) complemented with brightness of stars in column Mag. range.}
\begin{tabular}{l c c c c c c c}\\
\hline
Name   & $R$  &  $r$     & $\alpha$  & $\delta$ & $P$ & VAR & Mag. range\\
       & [\%] & [arcsec] & [$\deg$]  & [$\deg$] & [d] &     & \\
\hline\hline\\[-3mm]
CI CrB                & 0.000 & 0.00 & 242.08654 & +28.20825 & 0.350076 & EW: & 12.66 (0.178) R1 \\
CSS\_J160820.7+281300 & 0.011 & 30.6 & 242.08658 & +28.21669 & 0.350038 & EW  & 13.10 (0.22) CV \\
CSS\_J160820.5+281301 & 0.013 & 31.8 & 242.08550 & +28.21706 & 0.35003  & EW  & 13.12 (0.19) CV \\
CSS\_J160820.5+281303 & 0.013 & 33.6 & 242.08567 & +28.21753 & 0.35003  & EW  & 13.16 (0.25) CV \\
\hline\\
\end{tabular}\label{table2}
\end{table}

\begin{figure}[t!]
\centering
\includegraphics[width=14cm]{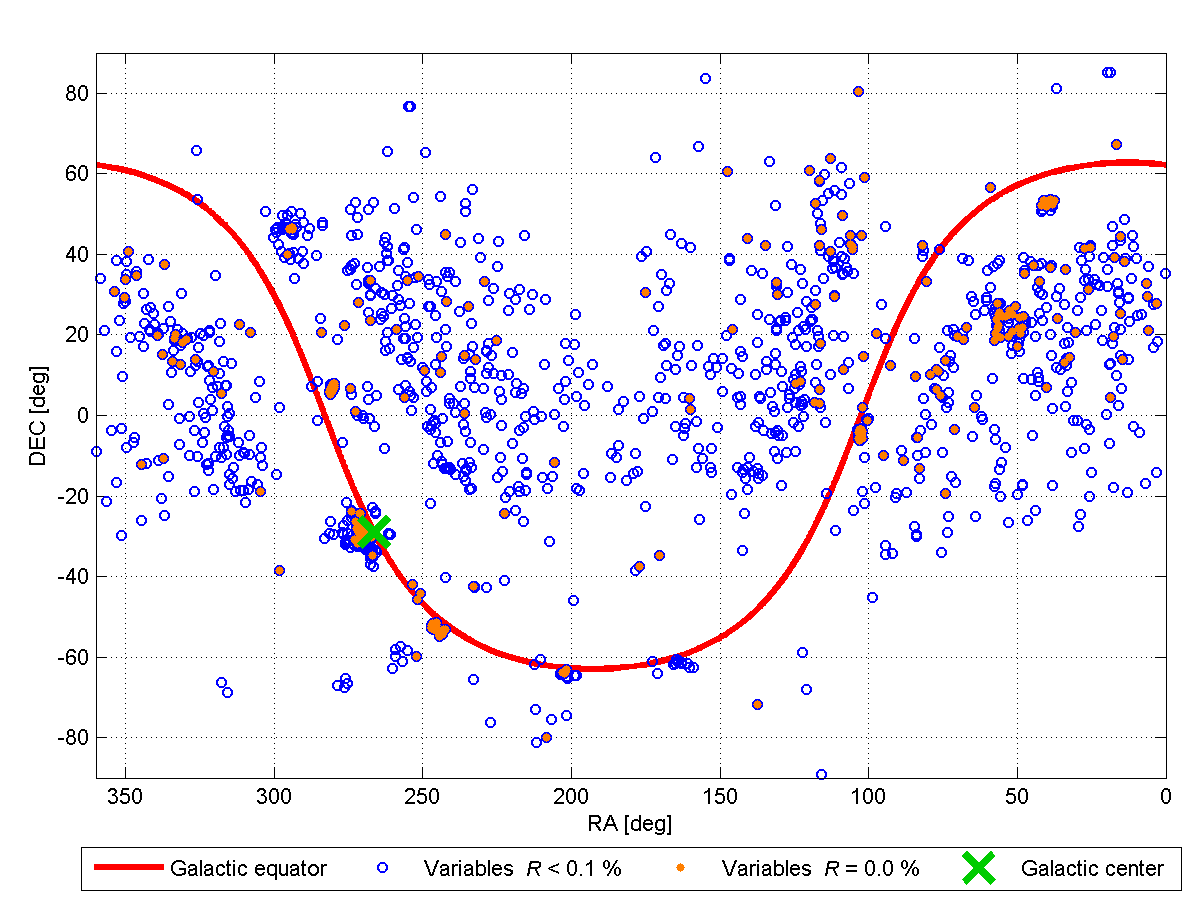} 
\caption{A distribution of possible misclassified variables with mutual distance $r<1$\,arcmin and period differences $R<0.1$\,\% and $R=0.0$\,\% in equatorial coordinates. Galactic equator is shown by the red line, the green cross shows the position of the Galactic center. Note the condensation around the Galactic center with stars located in the Galactic bulge.}
\label{fig2}
\end{figure}

The sky position of positive identifications with $R<0.1$\,\% and $R=0.0$\,\% are shown in Fig.~\ref{fig2}. As can be seen, the distribution is more or less random and proposed duplicates do not follow the Milky Way as was expected. However, the most of duplicates are located in the Galactic bulge. This should not be surprising since this region contains a huge number of very close stars and is intensively observed by MACHO and OGLE surveys \citep[e.g.][]{bennet1991,udalski1992}.

\begin{figure}[t!]
\centering
\includegraphics[width=14cm]{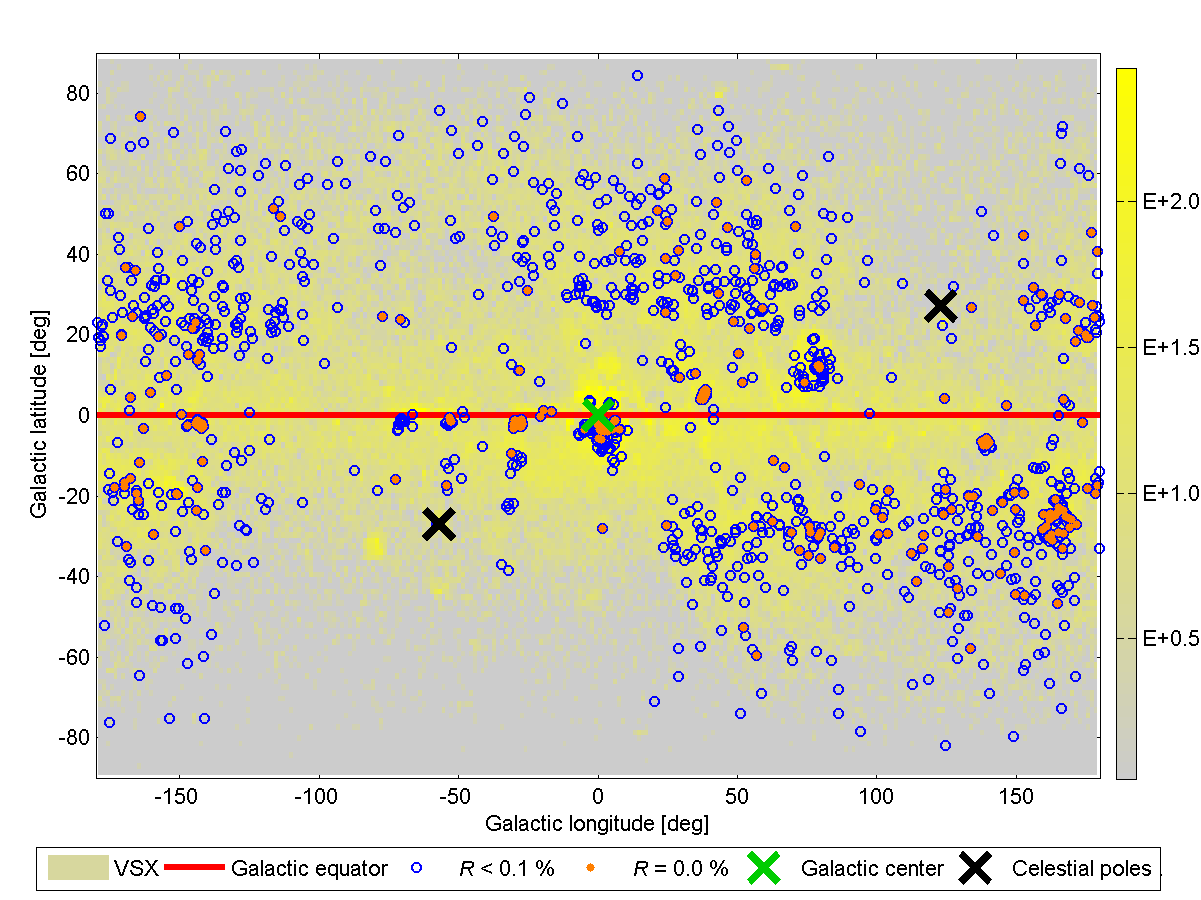} 
\caption{The same distribution as in Fig.~\ref{fig2} in galactic coordinates, together with all variables in VSX (density of stars in logarithmic scale, gray-yellow) and Celestial poles (black crosses). The lack of duplicates along Milky Way and around Celestial and Galactic poles is apparent.}
\label{fig3}
\end{figure}

\begin{figure}[t!]
\centering
\includegraphics[width=14cm]{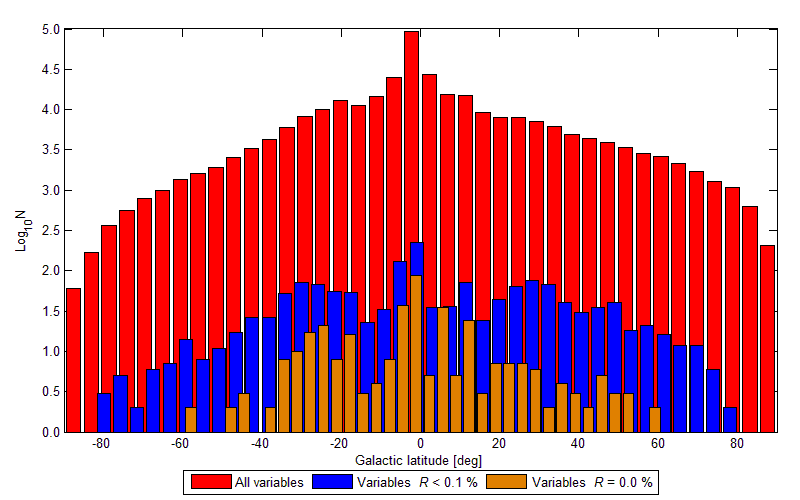} 
\caption{A distribution of variable stars in VSX and our candidates in logarithmic scale as a function of galactic latitude.}
\label{fig4}
\end{figure}

A closer investigation of Fig.~\ref{fig2} and \ref{fig3}, which shows distribution in galactic coordinates, reveals a lack of potential duplicates along Milky Way. This is very surprising because it suggests that all-sky surveys identify variables in crowded fields with higher efficiency. We do not have any explanation for this finding because VSX contains variables in this area\footnote{222824 stars from VSX catalogue (68.55\,\%) are in area within galactic latitudes $-20^{\circ}$ and $+20^{\circ}$.} which is seen in Fig.~\ref{fig3} and \ref{fig4}. Several sky-surveys such as ASAS, NSV and NSVS cover this area, another projects (e.g. CSS) avoid the Galactic plane (see Appendix A). Except for the Milky Way there is lack of duplicates around Celestial and Galactic poles. We assume that this fact corresponds with decreasing number of stars towards the poles.

Our analysis also showed that more than half of the pairs with $R = 0.0$\,\% were identified by the same project (the same acronym in designations). The worst results were found for Catalina Sky Survey with 73 pairs which have both names with CSS acronyms and BEST projects \citep[e.g.,][]{kabath2009a,kabath2009b} with 71 pairs (numbers for other projects are in Tab.~\ref{table3}). Special case above all is EROS2 project \citep{derue2002}, in which all 51 pairs with EROS2 acronym have the same period $P>60$\,d. This 60\,d corresponds to the length of timebase and is defined as a lower limit for periods.

\begin{table}[h]
\setlength{\tabcolsep}{4pt}
\centering
\caption{Number of pairs with $R = 0.0$\,\% (almost certain duplicities) with both names from the same project.}
\begin{tabular}{l c c c c c c c c c}\\
\hline
Project & ASAS	& BEST  & BOKS	& CSS	& EROS2	& HAT	& MACHO & NSVS 	& OGLE\\
\hline\hline\\[-3mm]
Number	& 18 	& 71 	& 20	& 73	& (51)	& 23	& 2		& 4		& 18\\
\hline\\
\end{tabular}\label{table3}
\end{table}

Although the majority of stars in our sample should reflect the real number of duplicates, we cannot exclude the possibility that some of given objects are real, independent variables. In addition, several objects are known to create a wide binary pairs with period in ratio of small integer numbers \citep[e.g. BV and BW Dra with period ratio about 6:5, e.g.][]{batten1965}. Unresolved quadruple systems like CzeV343 \citep[double eclipsing binary with period ratio of 3:2,][]{cagas2012} could also bring some additional bias in our results when the ratio is 1:1 or 1:2. 

\newpage
An interesting situation brings the first pair of stars in Tab.~\ref{table1}. GM And is an RRab star with a pulsation period of 0.7067585\,d and V0467 And is a binary system of EW type with an eclipsing period of 0.3534\,d. Their angular distance is only about 38\,arcsec. When we used double value of period for V0467 And, their relative difference between periods is $R\sim0.0059$\,\% and thus well bellow our criterion $R<0.1$\,\%. The both variables have well-known types and periods -- GM And \citep{schmidt1993}, V0467 And \citep{maintz2008}.

V2660~Oph (EA, $P_{\rm eclipse}=1.2061988$\,d) \citet{haussler2005} and EP~Ser (RRab, $P_{\rm puls} = 0.6032100$\,d) \citet{haussler2006}, with $r = 58.46$\,arcsec and $R = 0.018$\,\%, are another example of close stars which are definitely different objects with close periods. The probability of two close stars having the same period is low but non-zero (even in sparse fields).

In addition, we found three Kepler objects which should be excluded from the main VSX catalogue (Tab.~\ref{table4}). \citet{coughlin2014} already realized that KID~10342065, KID~10342041 and KIC 10407221 are false positives identified as EA binaries, which is caused by very bright EA star V2083 Cyg (about 7 mag). Unfortunately, physical parameters of false binary systems KID~10342065 and KID~10342041 have been already determined based on light curve analysis by \citet{slawson2011}.  A huge influence of this bright star can be illustrated on KIC 10407221 which is 15.6\,mag star in a distance of 4.29\,arcmin. The information about brightness of stars could therefore not be used as additional criterion for duplicity.

\begin{table}[h!]
\setlength{\tabcolsep}{4pt}
\centering
\caption{Nearby objects of V2083~Cyg. The columns are identical as in Tab.~\ref{table2} ($R$ and $r$ parameters are relative to V2083~Cyg).}
\begin{tabular}{l c c c c c c c}\\
\hline
Name   & $R$  &  $r$     & $\alpha$   & $\delta$ & $P$ & VAR & Mag. range\\
       & [\%] & [arcsec] & [$\deg$]   & [$\deg$] & [d] &     & \\
\hline\hline\\[-3mm]
V2083~Cyg             & 0.000 & 0.00  & 292.81817 & +47.48117 & 1.86742  & EA & 6.94 -- 7.18 Hp \\
KID 10342065          & 0.014 & 59.1  & 292.84213 & +47.48383 & 0.933837 & EA & 13.964 (0.002) Kp \\
KID 10342041          & 0.007 & 63.0  & 292.83300 & +47.49553 & 0.933771 & EA & 14.658 -- ? Kp \\
KIC 10407221          & 0.002 & 257.4 & 292.84448 & +47.55042 & 0.933728 & EA & 15.647 -- ? Kp \\
\hline\\
\end{tabular}\label{table4}
\end{table}

\newpage
Finally we noticed another bias present in the VSX catalogue. Some long-period variables have unrealistically accurate periods (five and more decimal places)\footnote{For example, period accuracy of stars with period longer than 100\,days would be maximally 2 or 3 decimal places.}. Furthermore we identified several hundreds stars with periods longer than 300 days (especially from ASAS survey) spread out over the whole sky, which have exactly the same period. The extreme examples are 18 ASAS objects with period of 450.819672 d. These stars are probably not duplicates, but are very likely badly processed. The list with such objects is given on-line as a supporting information (file is sorted in VSX format).

Our study should serve as a warning for observers interested in objects we identified because they could have false identifications or be non-variable objects at all. 

\OEJVacknowledgements{This research has made use of the International Variable Star Index (VSX) database, operated at AAVSO, Cambridge, Massachusetts, USA. This work was supported by the grants MUNI/A/1110/2014 and LH14300. Our special thanks belong to the people who take care about the VSX. We thank referees for helpful comments and suggestions which improved our manuscript.}

\newpage
\section*{Appendix A}

\begin{figure}[ht]
\centering
\includegraphics[width=9.5cm]{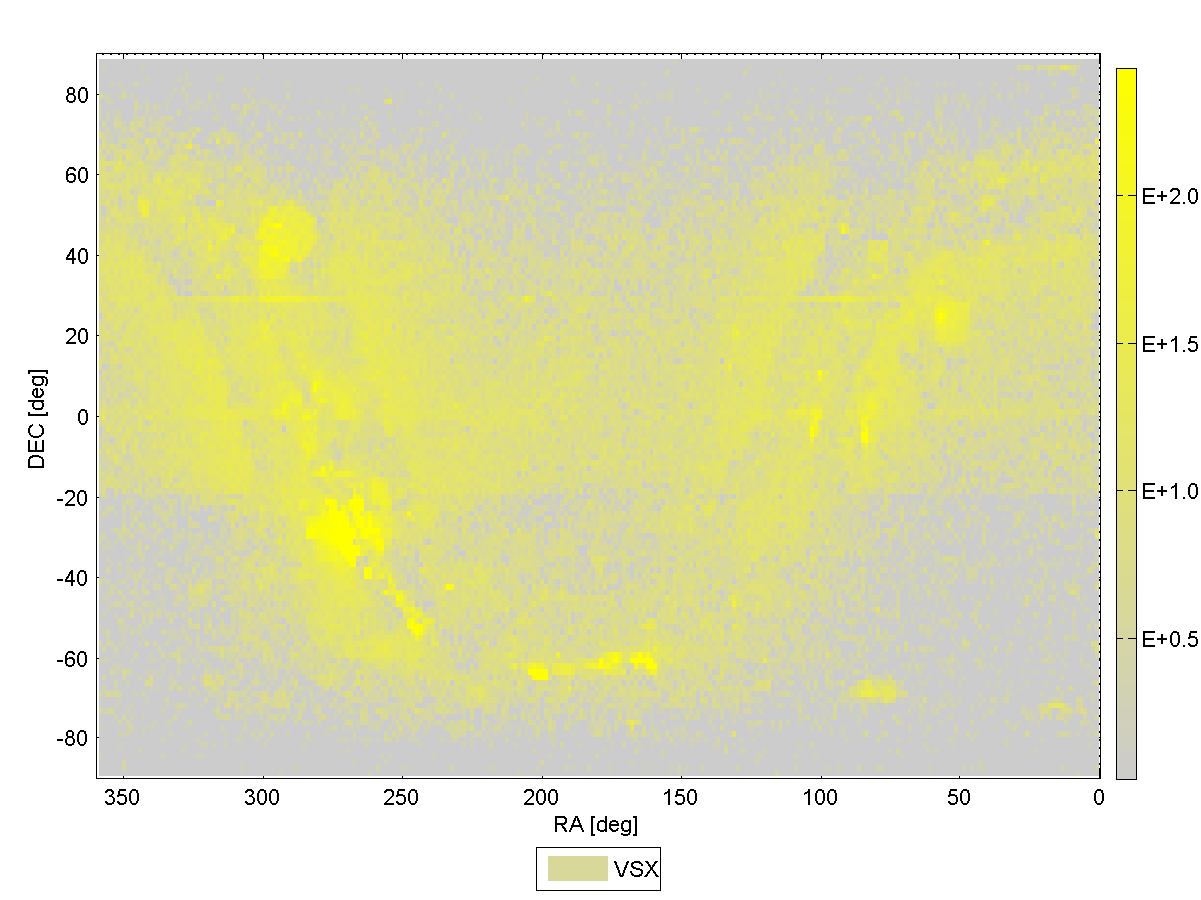} 
\caption{The distribution of 325061 variable stars from VSX in equatorial coordinates.}
\label{fig5}
\end{figure}

\begin{figure}[hb]
\centering
\includegraphics[width=9.5cm]{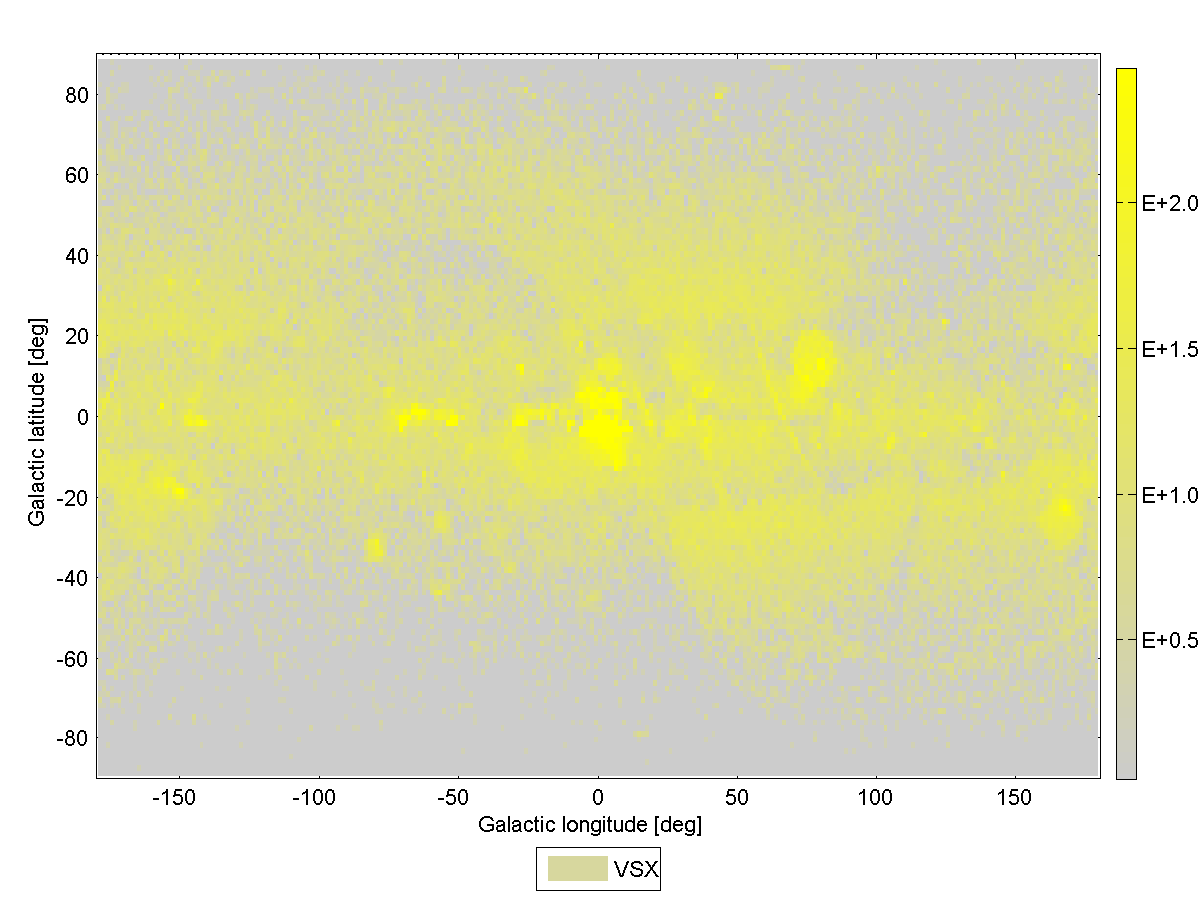} 
\caption{The distribution of 325061 variable stars from VSX in galactic coordinates.}
\label{fig6}
\end{figure}

\begin{figure}[hb]
\centering
\includegraphics[width=11cm]{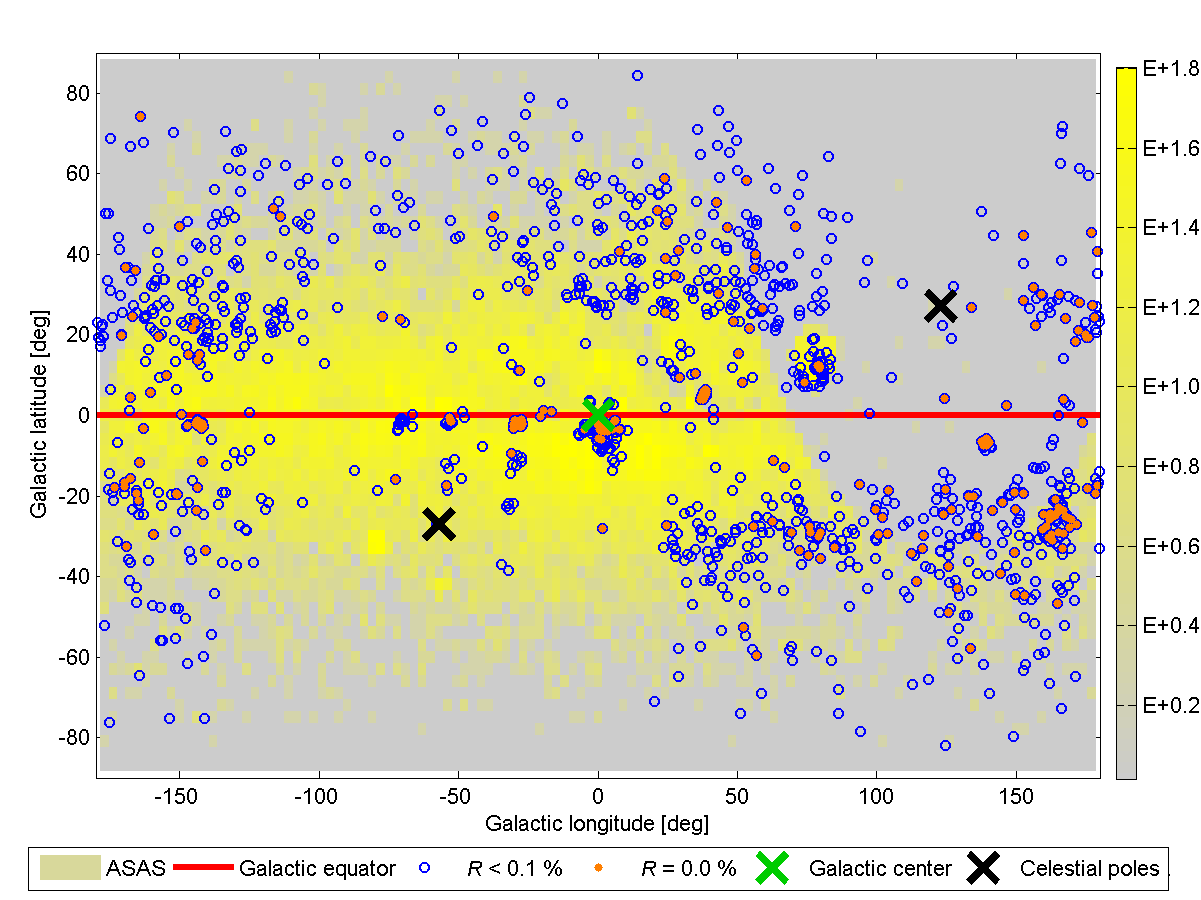} 
\caption{The distribution of 44846 variable stars with ASAS names (gray-yellow) in galactic coordinates.}
\label{fig7}
\end{figure}

\begin{figure}[ht]
\centering
\includegraphics[width=11cm]{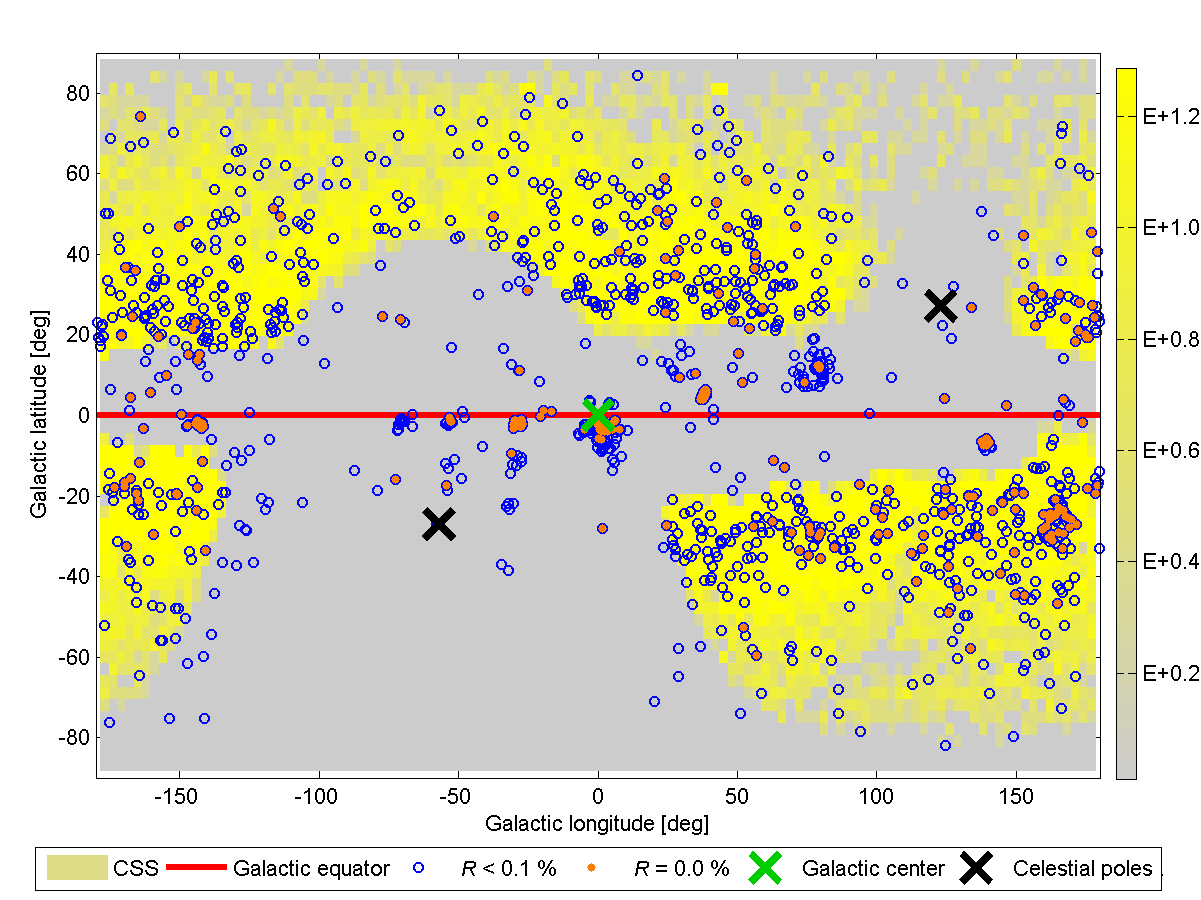} 
\caption{The distribution of 49834 variable stars with CSS names (gray-yellow) in galactic coordinates.}
\label{fig8}
\end{figure}

\begin{figure}[ht]
\centering
\includegraphics[width=11cm]{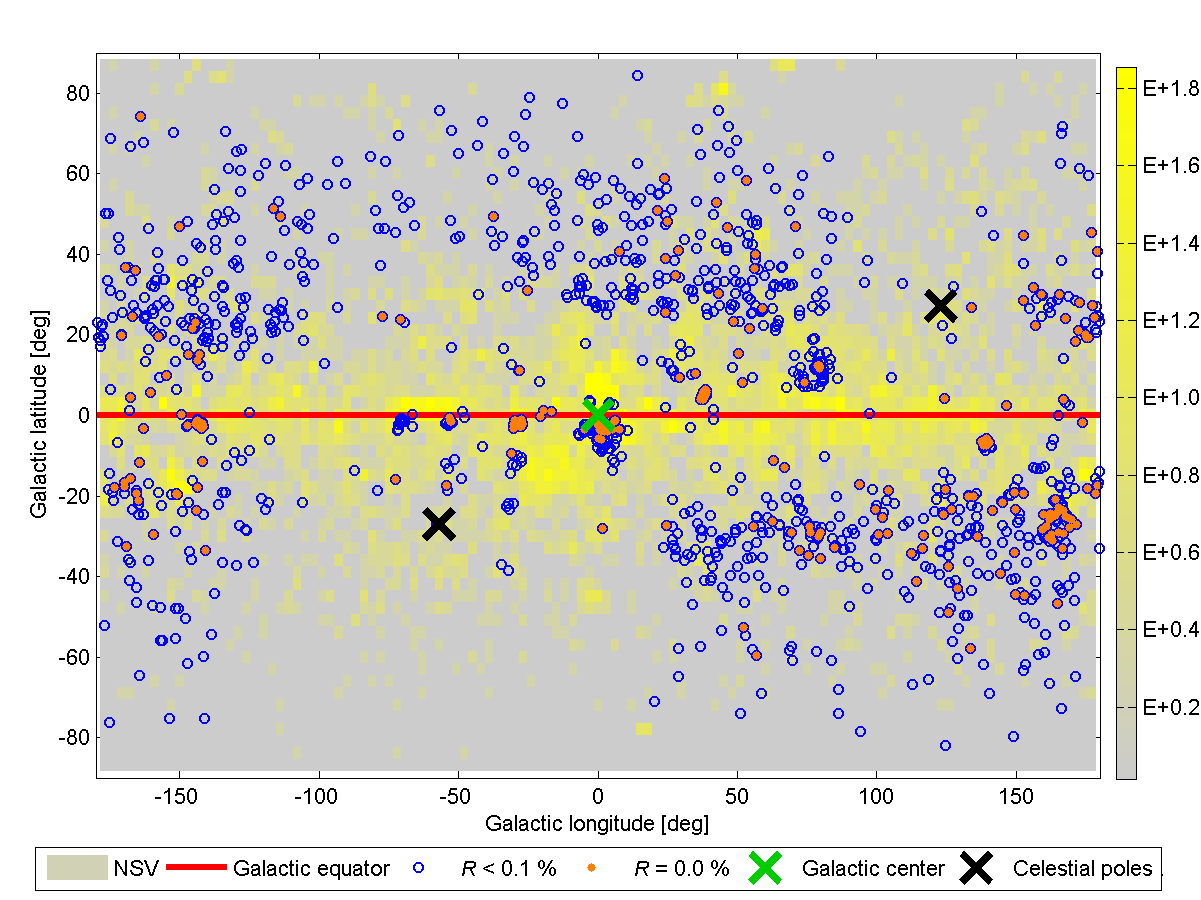} 
\caption{The distribution of 22177 variable stars with NSV names (gray-yellow) in galactic coordinates.}
\label{fig9}
\end{figure}

\begin{figure}[hb]
\centering
\includegraphics[width=11cm]{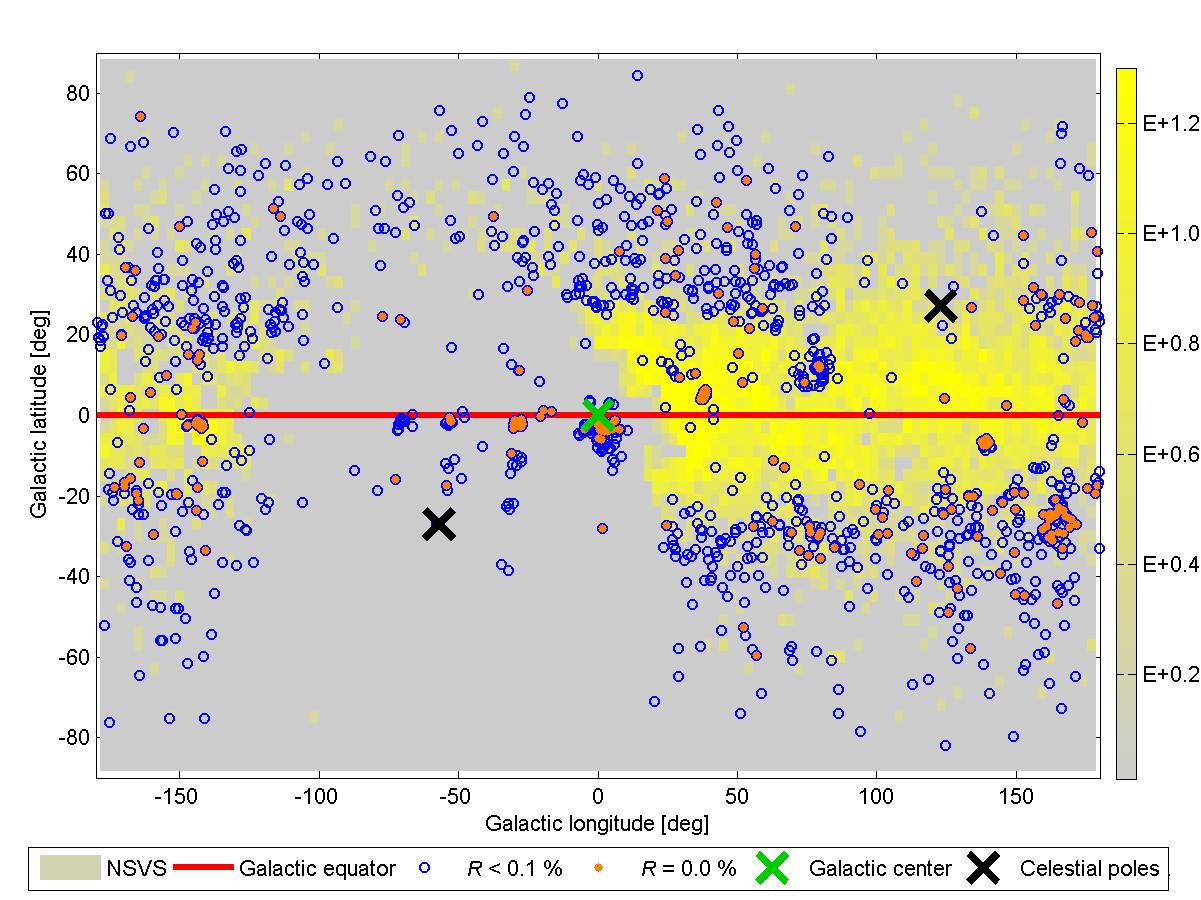} 
\caption{The distribution of 13058 variable stars with NSVS names (gray-yellow) in galactic coordinates.}
\label{fig10}
\end{figure}

\end{document}